\documentclass[prd,aps,a4paper,nofootinbib,twocolumn,showpacs,showkeys]{revtex4}  

\newif\ifusesec
\usesectrue  
   
\usepackage{graphicx} 
\usepackage{mathrsfs}
\usepackage{latexsym,amsmath,amsfonts,amssymb}
\usepackage{multirow}

\newcommand{\beq}{\begin{equation}}
\newcommand{\eeq}{\end{equation}}
\begin{document}

\title{Hyperbolic scattering of spinning particles by a Kerr black hole}

\author{Donato \surname{Bini}$^{1,2}$}
\author{Andrea \surname{Geralico}$^{1,2}$}
\author{Justin Vines${}^3$}

\affiliation{
$^1$Istituto per le Applicazioni del Calcolo ``M. Picone'', CNR, I-00185 Rome, Italy\\
$^2$ICRANet, Piazza della Repubblica 10, I-65122 Pescara, Italy \\
$^3$Max Planck Institute for Gravitational Physics (Albert Einstein Institute),
Am M\"uhlenberg 1, 14476 Potsdam-Golm, Germany, EU
}

\date{\today}

\begin{abstract}
We investigate the scattering of a spinning test particle by a Kerr black hole within the Mathisson-Papapetrou-Dixon model to linear order in spin.
The particle's spin and orbital angular momentum are taken to be aligned with the black hole's spin. Both the particle's mass and spin length are assumed to be small in comparison with the characteristic length scale of the background curvature, in order to avoid backreaction effects.
We analytically compute the modifications due to the particle's spin to the scattering angle, the periastron shift, and the condition for capture by the black hole, extending previous results valid for the nonrotating Schwarzschild background. Finally, we  discuss how to generalize the present analysis beyond the linear approximation in spin, including spin-squared corrections in the case of a black-hole-like quadrupolar structure for the extended test body.
\end{abstract}

\pacs{04.20.Cv}
\keywords{Kerr black hole; spinning bodies; scattering process.}

\maketitle

\section{Introduction}

The scenario of a black hole which scatters matter, radiation and even other black holes is especially attractive in view of the possibility to consider these events as sources of gravitational waves, potentially detectable by current detectors such as LIGO and VIRGO \cite{ligo,virgo,Abbott:2016blz} or future detectors.
While much is known for \lq\lq elliptic encounters," or bound motion, namely the spiraling of one body around another up to the merging of the two bodies into a single one, the case of \lq\lq hyperbolic encounters," or unbound motion, has been poorly investigated in the literature.
In fact, up to now the back-reaction on a Kerr background of a small particle moving on an equatorial hyperbolic-like orbit has not yet been analytically computed.
The first attempt concerning a spinless particle in a Schwarzschild background has been presented only very recently in Ref. \cite{Hopper:2017qus}. Our  knowledge of the main features of the process mostly relies on the post-Newtonian approximation (at low orders; see, e.g., Refs. \cite{Damour:2014afa,Damour:2016gwp,Bini:2017wfr} and references therein) or numerical relativity simulations \cite{Shibata:2008rq,Sperhake:2008ga,Sperhake:2009jz,Sperhake:2012me}.
This lack of information is due to the computational difficulties associated with the problem, like the fact that the hyperbolic gravitational wave spectrum is a continuum, in contrast to the quasi-circular case, where it is mainly monochromatic. Furthermore, apart from the direct relevance to actual hyperbolic motions, the study of unbound motion is also relevant for the analysis of bound motion. In fact, for instance, computing the scattering angle for a hyperbolic encounter can encode gauge-invariant information characterizing both unbound and bound motion (see, e.g., Ref. \cite{Damour:2009sm}, where the gauge-invariant \lq\lq effective-one-body'' function $\rho(x)$ was introduced in the discussion of small eccentricity motion).

A necessary pre-requisite to the study of back-reaction effects due to gravitational wave emission is the complete knowledge of the conservative dynamics of a hyperbolic encounter of a test particle with the (spinning) black hole.  When the particle's internal structure is negligible, it moves along a hyperbolic-like geodesic orbit around the black hole. 
We study here the problem of the scattering of a particle endowed with spin by a Kerr black hole in the framework of the Mathisson-Papapetrou-Dixon (MPD) model \cite{Mathisson:1937zz,Papapetrou:1951pa,Dixon:1970zza}, generalizing a previous work in the nonrotating Schwarzschild background spacetime \cite{Bini:2017ldh}.
The motion is non-geodesic due to the presence of a spin-curvature coupling force.
The two bodies are assumed to have aligned/antialigned spins and the equatorial plane is chosen as the orbital plane.
According to the MPD model the extended body is treated as a test body, i.e., backreaction effects on the background metric are neglected, even if it has an internal structure described by its spin.
The MPD equations of motion can be solved analytically in terms of elliptic integrals, allowing to explicitly compute the corrections to first-order in spin to the scattering angle, i.e., the most natural gauge-invariant and physical observable associated with the scattering process.
We also determine the shift of the periastron position and the modification due to spin to the critical impact parameter for capture by the hole with respect to the well known geodesic case in both ultrarelativistic and non-relativistic regimes.
Finally, we show how to extend the present analysis to take into account spin-squared effects as well by suitably modifying the MPD set of equations, qualitatively discussing some features of the second-order-in-spin corrections to the orbital motion in comparison with the linear-in-spin ones, in the simplest case of an extended body endowed with a \lq\lq black-hole-like'' quadrupolar structure.

We use geometrical units and assume that Greek indices run from $0$ to $3$, whereas Latin indices from $1$ to $3$.

\section{Equatorial plane motion in a Kerr spacetime}

The motion of a spinning test particle in any given gravitational background is described by the MPD equations \cite{Mathisson:1937zz,Papapetrou:1951pa,Dixon:1970zza}
\begin{eqnarray}
\label{papcoreqs1}
\frac{ DP^{\mu}}{d \tau} & = &
- \frac12 \, R^\mu{}_{\nu \alpha \beta} \, U^\nu \, S^{\alpha \beta}
\equiv  F^\mu_{\rm (spin)}\,,
\\
\label{papcoreqs2}
\frac{ DS^{\mu\nu}}{d \tau} & = & 
2 \, P^{[\mu}U^{\nu]} \,.
\end{eqnarray}
Here $P^{\mu}\equiv m u^\mu$ (with $u \cdot u \equiv u^\mu u_\mu = -1$) is the total 4-momentum of the body with mass $m$ and unit timelke four-velocity $u^\mu$; $S^{\mu\nu}$ is the (antisymmetric) spin tensor; and $U^\mu=d x^\mu/d\tau$ is the timelike unit 4-velocity vector tangent to the body's ``center-of-mass worldline,'' parametrized by the proper time $\tau$ (with parametric equations $x^\mu=x^\mu(\tau)$), selected to perform the multipole reduction which is at the basis of the MPD approach.

Self-consistency of the model requires that some additional conditions be imposed to the determine reference world line, as Eqs.~(\ref{papcoreqs1}) and (\ref{papcoreqs2}) represent only 10 equations for the 14 degrees of freedom $x^\mu(\tau)$, $P^\mu(\tau)$ and $S^{\mu\nu}(\tau)$. Here we shall use the Tulczyjew-Dixon conditions~\cite{Dixon:1970zza,tulc59}, which read
\beq
\label{tulczconds}
S^{\mu\nu}u{}_\nu=0\,. 
\eeq
The spin tensor can then be fully represented by the spatial vector (with respect to $u$)
\beq
\label{svec}
S(u)^\alpha=\frac12
\eta(u)^\alpha{}_{\beta\gamma}S^{\beta\gamma}
\,,\qquad S^{\mu\nu}=\eta(u)^{\mu\nu}{}_\alpha S(u)^\alpha\,,
\eeq 
where we have denoted by $\eta(u)_{\alpha\beta\gamma}=\eta_{\mu\alpha\beta\gamma}u^\mu$ the
spatial unit volume 3-form (with respect to $u$) built from the unit volume 4-form
$\eta_{\alpha\beta\gamma\delta}=\sqrt{-g}\, \epsilon_{\alpha\beta\gamma\delta}$ which orients the spacetime,
with $\epsilon_{\alpha\beta\gamma\delta}$ ($\epsilon_{0123}=1$) being the
Levi-Civita alternating symbol and $g$ the determinant of the metric. 

It is also useful to introduce the signed
magnitude $s$ of the spin vector
\beq
\label{sinv}
s^2=S(u)^\beta S(u)_\beta = \frac12 S_{\mu\nu}S^{\mu\nu}
\,,
\eeq
which is a constant of motion.
Implicit in the MPD model is the condition that the length scale $|s|/m$ naturally associated with the spin should be very small compared to the one associated with the background curvature (say $M$), in order to neglect back reaction effects, namely $|\hat s|\equiv {|s|}/{mM}\ll 1$.
This  condition leads to a simplified set of linearized evolution equations around the geodesic motion. 
In fact, the total 4-momentum $P$ of the particle is still aligned with $U$ in this limit, i.e., $P^{\mu}= mU^\mu+O(S^2)$, the mass $m$ of the particle remaining constant along the path. Furthermore, Eq.~(\ref{papcoreqs2}) becomes ${ DS^{\mu\nu}}/{d \tau}=O(S^2)$, implying that the spin vector is parallel transported along $U$. 

Finally, when the background spacetime has Killing vectors, there are conserved quantities along the motion \cite{ehlers77}, which can be used to simplify the equations of motion. In the case of stationary axisymmetric spacetimes with coordinates adapted to the spacetime symmetries, they are the energy $E$ and the total angular momentum $J$ associated with the timelike and azimuthal Killing vectors $\xi=\partial_t$ and $\eta=\partial_\phi$, respectively, and read
\beq
\label{totalenergy}
E =-P_t +\frac12 S^{\alpha\beta}\xi_{\alpha; \beta} \,,\qquad
J =P_\phi -\frac12 S^{\alpha\beta}\eta_{\alpha; \beta}\,.
\eeq

\subsection{Kerr spacetime and ZAMO adapted frame}

Let us consider the Kerr spacetime, whose metric written in standard Boyer-Lindquist coordinates is given by
\begin{eqnarray}
ds^2 &=& -\left(1-\frac{2Mr}{\Sigma}\right)dt^2 
-\frac{4aMr}{\Sigma}\sin^2\theta dtd\phi+ \frac{\Sigma}{\Delta}dr^2\nonumber\\
&&+\Sigma d\theta^2+\frac{A}{\Sigma}\sin^2 \theta d\phi^2\,,
\end{eqnarray}
with $\Delta=r^2-2Mr+a^2$, $\Sigma=r^2+a^2\cos^2\theta$ and $A=(r^2+a^2)^2-\Delta a^2\sin^2\theta$.
Here $a$ and $M$ denote the specific angular momentum and the total mass of the spacetime solution, so that the quantity $\hat a=a/M$ is dimensionless.
The inner and outer horizons are located at $r_\pm=M\pm\sqrt{M^2-a^2}$.

Let us introduce the zero angular momentum observer (ZAMO) family of fiducial observers, with 4-velocity
\beq
\label{n}
n=N^{-1}(\partial_t-N^{\phi}\partial_\phi)\,,
\eeq
where $N=(-g^{tt})^{-1/2}=\sqrt{\Delta\Sigma/A}$ and $N^{\phi}=g_{t\phi}/g_{\phi\phi}=-2aMr/A$ are the lapse and shift functions, respectively. 
The ZAMO adapted form of the metric then reads
\beq
d s^2 = -N^2d t^2 +g_{rr}d r^2+g_{\theta\theta}d \theta^2+g_{\phi\phi}(d \phi+N^\phi d t)^2
\,.
\eeq
A suitable spatial orthonormal frame
adapted to ZAMOs is given by
\begin{eqnarray}
\label{ZAMO-frame}
e_{\hat t}&=&n\,, \quad 
e_{\hat r}=\frac1{\sqrt{g_{rr}}}\partial_r\equiv\partial_{\hat r}\,,\nonumber\\
e_{\hat \theta} &=&
\frac1{\sqrt{g_{\theta \theta }}}\partial_\theta\equiv\partial_{\hat\theta}\,, \quad
e_{\hat \phi}=\frac1{\sqrt{g_{\phi \phi }}}\partial_\phi\equiv\partial_{\hat\phi}\,.
\end{eqnarray}
The nonzero ZAMO kinematical quantities (i.e., acceleration $a(n)=\nabla_n n$ and expansion vector $\theta_{\hat \phi}(n)^\alpha=\theta(n)^\alpha{}_\beta\,{e_{\hat\phi}}^\beta$) all belong to the $r$-$\theta$ 2-plane of the tangent space \cite{Jantzen:1992rg,Bini:1997ea,Bini:1997eb,Bini:1999wn}, i.e.,
\begin{eqnarray}
\label{accexp}
a(n) & = & a(n)^{\hat r} e_{\hat r} + a(n)^{\hat\theta} e_{\hat\theta}\nonumber\\
&\equiv&\partial_{\hat r}(\ln N) e_{\hat r} + \partial_{\hat\theta}(\ln N)  e_{\hat\theta}
\,,
\nonumber\\
\theta_{\hat\phi}(n) & = & \theta_{\hat\phi}(n)^{\hat r}e_{\hat r} 
+ \theta_{\hat\phi}(n)^{\hat\theta}e_{\hat \theta}\nonumber\\ 
&\equiv& -\frac{\sqrt{g_{\phi\phi}}}{2N}\,(\partial_{\hat r} N^\phi e_{\hat r} 
+ \partial_{\hat\theta} N^\phi e_{\hat \theta})\,.
\end{eqnarray}
On the equatorial plane ($\theta=\pi/2$) the only nonvanishing components are given by
\begin{eqnarray}
a(n)^{\hat r} &=& \frac{M}{r^2\sqrt{\Delta}}
\frac{(r^2+a^2)^2-4a^2Mr}{r^3+a^2r+2a^2M}\,,\nonumber\\
\theta_{\hat\phi}(n)^{\hat r}&=& -\frac{aM(3r^2+a^2)}{r^2(r^3+a^2r+2a^2M)}\,.
\end{eqnarray}

\subsection{Spinning particle motion on the equatorial plane}

Let us consider a spinning particle moving on the equatorial plane of a Kerr spacetime with the spin vector aligned with the spacetime rotation axis, i.e.,
\beq
\label{SUdef}
S(U)=-se_{\hat \theta}\,. 
\eeq
Its 4-velocity $U$ decomposed with respect to the ZAMOs is
\beq
\label{Uzamo}
U=\gamma(U,n) [n+ \nu(U,n)]\,,\quad 
\eeq
where 
\beq
\nu(U,n)\equiv \nu(U,n)^{\hat r}e_{\hat r}+\nu(U,n)^{\hat \phi}e_{\hat \phi}\,,
\eeq
and 
$\gamma(U,n)=1/\sqrt{1-||\nu(U,n)||^2}$ is the associated Lorentz factor.
Hereafter we will use the abbreviated notation $\gamma(U,n)\equiv\gamma$ and $\nu(U,n)^{\hat a}\equiv\nu^{\hat a}$.

The equations of motion (\ref{papcoreqs1}) then imply
\begin{eqnarray}
\label{setfin}
\frac{d\nu^{\hat r}}{d \tau}&=&
-\gamma\left\{(\nu^{\hat \phi})^2k_{\rm (lie)}\right. \nonumber\\
&+& \left. (1-(\nu^{\hat r})^2)\left[a(n)^{\hat r}+2\theta_{\hat\phi}(n)^{\hat r}\nu^{\hat \phi}\right]\right\}\nonumber\\
&+&
\frac1{m\gamma}\left[(1-(\nu^{\hat r})^2)F_{\rm(spin)}^{\hat r}-\nu^{\hat r}\nu^{\hat \phi}F_{\rm(spin)}^{\hat \phi}\right]
\,,\nonumber\\
\frac{d \nu^{\hat \phi}}{d \tau}&=&
\gamma\nu^{\hat r}\nu^{\hat \phi}\left[k_{\rm (lie)}+a(n)^{\hat r}+2\theta_{\hat\phi}(n)^{\hat r}\nu^{\hat \phi}\right]\nonumber\\
&+&
\frac1{m\gamma}\left[-\nu^{\hat r}\nu^{\hat \phi}F_{\rm(spin)}^{\hat r}+(1-(\nu^{\hat \phi})^2)F_{\rm(spin)}^{\hat \phi}\right]\,,\nonumber\\
\end{eqnarray}
where
\beq
k_{\rm (lie)}=-\frac{(r^3-a^2M)\sqrt{\Delta}}{r^2(r^3+a^2r+2a^2M)}
\eeq
denotes the Lie relative curvature \cite{Bini:1997ea,Bini:1997eb} evaluated at $\theta=\pi/2$, defined by $k_{\rm (lie)}=-\partial_{\hat r}\ln\sqrt{g_{\phi\phi}}$.
The frame components of the spin force with respect to ZAMOs are given by
\begin{eqnarray}
F_{\rm(spin)}^{\hat r}&=&mM\hat s\gamma^2\left[(E_{\hat \theta \hat \theta}- E_{\hat r \hat r})\nu^{\hat \phi}+(1+(\nu^{\hat \phi})^2)H_{\hat r \hat \theta}\right]
\,,\nonumber\\
F_{\rm(spin)}^{\hat \phi}&=&-mM\hat s\gamma^2\left[(E_{\hat \theta \hat \theta}-E_{\hat \phi \hat \phi})\nu^{\hat r}+\nu^{\hat r}\nu^{\hat \phi}H_{\hat r \hat \theta}\right]
\,,
\end{eqnarray}
where
\begin{eqnarray} 
\label{E_H}
E_{\hat r \hat r}&=& -\frac{M(2 r^4+5 r^2 a^2-2 a^2 M r+3 a^4)}{r^4 (r^3+a^2r+2 a^2 M)}\,,\nonumber\\
E_{\hat \theta \hat \theta}&=&-E_{\hat \phi \hat \phi}- E_{\hat r \hat r}\,,\nonumber\\
E_{\hat \phi \hat \phi}&=&\frac{M}{r^3}\,,\nonumber\\
H_{\hat r \hat \theta}&=&  -\frac{3 M a (r^2+a^2) \sqrt{\Delta}}{r^4 (r^3+a^2 r+2 a^2 M)}\,,
\end{eqnarray}
are the nontrivial components of the electric and magnetic parts of the Riemann tensor with respect to ZAMOs, and $\hat s\equiv s/mM$ is the dimensionless signed spin magnitude.
The remaining nonvanishing component $F_{\rm(spin)}^{\hat t}$ follows from the condition $F_{\rm(spin)}\cdot U = 0$.
Finally, Eq.~(\ref{setfin}) must be coupled with the decomposition~(\ref{Uzamo}) of the 4-velocity $U=d x^\alpha/d\tau$, i.e.,
\begin{eqnarray}
\label{Ucompts}
\frac{dt}{d \tau} &=& \frac{\gamma}{N}\,,\nonumber\\
\frac{dr}{d \tau} &=& \frac{\gamma\nu^{\hat r}}{\sqrt{g_{rr}}}\,,\nonumber\\
\frac{d\phi}{d \tau} &=& \frac{\gamma}{\sqrt{g_{\phi\phi}}}
\left(\nu^{\hat \phi}-\frac{\sqrt{g_{\phi\phi}}N^\phi}{N}\right)\,,
\end{eqnarray}
providing the evolution of $t$, $r$ and $\phi$.

In order to solve the full set of equations (\ref{setfin}) and (\ref{Ucompts}) we take advantage of the existence of the conserved quantities (\ref{totalenergy}), which become
\begin{eqnarray}
\label{EandJ}
\hat E&=&N\gamma\left\{1-\frac{\sqrt{g_{\phi\phi}}N^\phi}{N}\nu^{\hat \phi}\right.\nonumber\\
&&
+M\hat s\left[a(n)^{\hat r}\nu^{\hat \phi}+\theta_{\hat\phi}(n)^{\hat r}\right. \nonumber\\
&& \left.\left. 
+\frac{\sqrt{g_{\phi\phi}}N^\phi}{N}\left(k_{\rm (lie)}+\theta_{\hat\phi}(n)^{\hat r}\nu^{\hat \phi}\right)\right]\right\}
\,,\nonumber\\
\hat J&=&\frac{\sqrt{g_{\phi\phi}}}{M}\gamma\left[\nu^{\hat \phi}-M\hat s\left(k_{\rm (lie)}+\theta_{\hat\phi}(n)^{\hat r}\nu^{\hat \phi}\right)\right]
\,,
\end{eqnarray}
where $\hat E\equiv E/m$ and $\hat J\equiv J/mM$ are dimensionless.
Eq. (\ref{EandJ}) thus provide two algebraic relations for the frame components $\nu^{\hat r}$ and $\nu^{\hat \phi}$ of the linear velocity, which once inserted in Eq. (\ref{Ucompts}) finally yield
\begin{widetext}
\begin{eqnarray} 
\label{finaleqs}
\frac{\Delta}{M^2} \frac{d t}{d \tau}&=&
{\hat E}\left(\frac{r^2}{M^2}+\hat a^2\right)+\frac{M}{r}[(2\hat a+3\hat s)\hat E-(2\hat a+\hat s)\hat J]-\frac{M^3}{r^3}\hat a^2\hat s(\hat J-\hat a\hat E)
\,, \nonumber\\ 
\left(\frac{d r}{d \tau}\right)^2&=&
{\hat E}^2-1+\frac{2M}{r}+\frac{M^2}{r^2}[\hat a^2({\hat E}^2-1)-{\hat J}({\hat J}-2{\hat E}\hat s)]\nonumber\\ 
&&
+\frac{2M^3}{r^3}(\hat J-\hat a\hat E)(\hat J-\hat a\hat E-3{\hat E}\hat s)
+\frac{2M^5}{r^5}\hat a\hat s(\hat J-\hat a\hat E)^2
\,, \nonumber\\ 
\frac{\Delta}{M}\frac{d\phi}{d \tau}&=&
{\hat J}-{\hat E}\hat s-\frac{2M}{r}(\hat J-\hat a\hat E-{\hat E}\hat s)-\frac{M^3}{r^3}\hat a\hat s(\hat J-\hat a\hat E)
\,,
\end{eqnarray}
\end{widetext}
to first order in spin.

Let us introduce a conical-like representation of the radial variable, i.e.,
\beq
\label{conica}
r=\frac{Mp}{1+e\cos \chi}\,,
\eeq
where both the semi-latus rectum $p$ and the  eccentricity $e\geq0$ are dimensionless parameters \cite{Chandrasekhar:1985kt}.
Bound orbits have $0\leq e<1$ and $0<\hat E<1$ and oscillate between a minimum radius $r_{\rm(per)}$ (periastron, $\chi=0$) and a maximum radius $r_{\rm(apo)}$ (apastron, $\chi=\pi$)
\beq
\label{periapodef}
r_{\rm(per)}=\frac{Mp}{1+e}\,,\qquad
r_{\rm(apo)}=\frac{Mp}{1-e}\,,
\eeq
corresponding to the extremal points of the radial motion, i.e., 
\beq
\label{periapo}
\frac{dr}{d\tau}\Big|_{r=r_{\rm(per)}}=0=\frac{dr}{d\tau}\Big|_{r=r_{\rm(apo)}}\,.
\eeq
Unbound (hyperbolic-like) orbits, instead, have eccentricity $e>1$ and energy parameter $\hat E>1$.
In particular, a typical scattering orbit starts far from the hole at radial infinity, reaches a minimum approach distance $r_{\rm(per)}$, and then comes back to radial infinity, corresponding to a scattering angle $\chi\in [-\chi_{\rm (max)}, \chi_{\rm (max)}]$, $\chi_{\rm (max)}=\arccos(-1/e)$ (see Eq. (\ref{conica})). In this case the apoastron does not exist anymore, as it corresponds to a negative value of the radial variable. However, the parametrization (\ref{periapodef}) can still be adopted.

The conditions (\ref{periapo}) allow one to express $\hat E$ and $\hat J$ in terms of $(p,e)$ as follows
\beq
\hat E =\hat E_0+\hat s\hat E_{\hat s}\,, \qquad
\hat J =\hat J_0+\hat s\hat J_{\hat s}\,, 
\eeq
to first order in spin.
The geodesic values are given by
\beq\label{Epex}
\hat E_0^2 =\frac{1}{p}\left[ \frac{(1-e^2)^2  \hat x_0^2}{p^2}+p-(1-e^2) \right] \,,
\eeq
where
\beq\label{x2sol}
\hat x_0^2 =\frac{-N\mp \sqrt{N^2-4CF}}{2F}\,, 
\eeq
with $F$, $N$ and $C$ dimensionless coefficients given by \cite{Glampedakis:2002ya}
\begin{eqnarray}
 F  &=&  \left(1-\frac{ 3+e^2 }{p} \right)^2-\frac{4\hat a^2(1-e^2)^2}{p^3}  \,,\nonumber\\
-\frac{N}{2}&=&(p-3-e^2)+\hat a^2\left(1 +\frac{1+3e^2}{p}\right)\,, \nonumber\\
C&=&(\hat a^2-p)^2\,,
\end{eqnarray}
and
\beq\label{discriminant}
N^2-4CF=\frac{16\hat a^2}{p^3}\{[p^2-2p+\hat a^2(1+e^2)]^2
-4e^2(p-\hat a^2)^2]\}\,.
\eeq
The upper (lower) sign corresponds to prograde (retrograde) motion.
The geodesic angular momentum is simply related to $\hat E_0$ and $\hat x_0$ by $\hat x_0=\hat J_0-\hat a\hat E_0$.
Formulas valid for retrograde orbits can be obtained from those for prograde orbits by $\hat a\to -\hat a$ and $\hat J_0 \to - \hat J_0$, under which $\hat x_0\to -\hat x_0$.

The first-order corrections to the energy and angular momentum are given by 
\begin{eqnarray}
\hat E_{\hat s}&=&
-\frac{\hat x_0}{2p^5}(1-e^2)^2[(3+e^2){\hat x_0}^2+p^2]
\,,\nonumber\\
\hat J_{\hat s}&=&
\hat E_0-\frac{\hat x_0}{2p^2}\hat E_0\hat J_0(3+e^2)-\frac{2\hat x_0}{p^3}\hat a(1-e^2)
\,.
\end{eqnarray}

\section{Hyperbolic-like motion}

After converting the radial equation into an equation for $\chi$ through Eq. (\ref{conica}), the azimuthal equation finally becomes
\begin{widetext}
\begin{eqnarray} 
\label{dphidchi}
\frac{d\phi}{d\chi}&=& u_p^{1/2}\frac{ \hat J_0 - 2 u_p \hat x_0 (1+  e\cos \chi) }{[1+u_p^2 \,\hat x_0^2 (e^2-2 e\cos \chi -3) ]^{1/2}
[1-2 u_p(1+ e\cos \chi) +\hat a^2 u_p^2(1+ e\cos \chi)^2  ]}\nonumber\\
&&
-\hat s u_p^{5/2}\hat x_0(\hat x_0\hat E_0+\hat a)\frac{e\cos\chi+3}{[1+u_p^2 \,\hat x_0^2 (e^2-2 e\cos \chi -3) ]^{3/2}}
\,,
\end{eqnarray}
\end{widetext}
to first order in spin, with $u_p=1/p$.
The solution of this equation can be obtained analytically in terms of elliptic functions as
\beq
\phi(\chi)=\phi_0(\chi)+\hat s \phi_{\hat s}(\chi)\,,
\eeq
where
\begin{widetext}
\begin{eqnarray}
\phi_0(\chi)&=&
\frac{\kappa}{2\hat a^2e^2u_p^2\sqrt{eu_p\hat x_0^2}(b_+-b_-)} \left\{
[\hat J_0-2u_p\hat x_0(1+eb_+)]\left[\Pi\left(\frac{\pi}{2}-\frac{\chi}{2},k_+,\kappa\right)-\Pi\left(k_+,\kappa\right)\right]\right.\nonumber\\
&&\left.
-[\hat J_0-2u_p\hat x_0(1+eb_-)]\left[\Pi\left(\frac{\pi}{2}-\frac{\chi}{2},k_-,\kappa\right)-\Pi\left(k_-,\kappa\right)\right]
\right\}
\,,\nonumber\\
\phi_{\hat s}(\chi)&=&
\frac{\kappa(\hat E_0\hat x_0+\hat a)}{2\sqrt{eu_p}\hat x_0^2} \left\{
K(\kappa)-F\left(\frac{\pi}{2}-\frac{\chi}{2},\kappa\right)\right.\nonumber\\
&&\left.
-\frac{C+6\hat x_0^2}{C-2e\hat x_0^2}  \left[ E(\kappa)-E\left(\frac{\pi}{2}-\frac{\chi}{2},\kappa\right)
+\kappa\sqrt{e}\hat x_0\frac{\sin\chi}{(A-2e\hat x_0^2\cos\chi)^{1/2}}
\right]
\right\}
\,,
\end{eqnarray}
\end{widetext}
with $\phi(0)=0$ and 
\begin{eqnarray} 
\kappa^2&=&\frac{4eu_p^2\hat x_0^2}{1+(e-1)(e+3)\hat x_0^2}\,,\nonumber\\
k_\pm&=&\frac{2}{1+b_\pm}\,,\nonumber\\
b_\pm&=&\frac{1-\hat a^2u_p\pm\sqrt{1-\hat a^2}}{\hat a^2eu_p}
\,, \nonumber\\
u_p^2C&=&1+u_p^2\hat x_0^2(e^2-3)\,.
\end{eqnarray}
Here $K(k)$ and $F(\varphi,k)$ and $E(k)$ and $E(\varphi,k)$ are the complete and incomplete elliptic integrals of the first kind and of the second kind, respectively, defined by
\begin{eqnarray}
F(\varphi,k)&=&\int_0^{\varphi}\frac{dz}{\sqrt{1-k^2\sin^2z}}\,,\quad
K(k)=F(\pi/2,k)
\,,\nonumber\\
E(\varphi,k)&=&\int_0^{\varphi}\sqrt{1-k^2\sin^2z}\,dz\,,\quad
E(k)=E(\pi/2,k)\,,\nonumber\\
\end{eqnarray} 
whereas
\begin{eqnarray}
\Pi(\phi,n,k)&=&\int_0^{\varphi}\frac{dz}{(1-n\sin^2z)\sqrt{1-k^2\sin^2z}}\,,\nonumber\\
\Pi(n,k)&=&\Pi(\pi/2,n,k)\,,
\end{eqnarray}
are the incomplete and complete elliptic integrals of the third kind, respectively~\cite{Gradshteyn}.

Unbound orbits which are not captured by the black hole start at an infinite radius at the azimuthal angle $\phi=\phi(-\chi_{\rm (max)})$, 
the radius decreases to its periastron value at $\phi=0$ and then returns back to infinite value at $\phi=\phi(\chi_{\rm (max)})$, undergoing a total increment of $\Delta\phi =\phi(\chi_{\rm (max)})-\phi(-\chi_{\rm (max)})$. 
This scattering process is symmetric with respect to the minimum approach ($\chi=0$) as in the case of a spinless particle, i.e., $\phi(-\chi_{\rm (max)})=-\phi(\chi_{\rm (max)})$, so that $\Delta\phi =2\phi(\chi_{\rm (max)})$, and the deflection angle from the original direction of the orbit is
\beq
\delta(u_p,e,\hat s)=\delta_0(u_p,e)+\hat s \delta_{\hat s}(u_p,e)\,,
\eeq
with $\delta_0(u_p,e)=2\phi_0(\chi _{\rm(max)})-\pi$ and $\delta_{\hat s}(u_p,e)=2\phi_{\hat s}(\chi_{\rm (max)})$.

\begin{figure}
\centering
\includegraphics[scale=0.4]{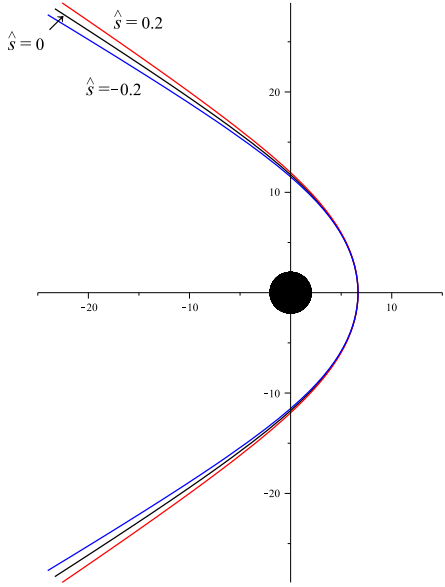}
\caption{The hyperbolic-like orbits of a spinning test particle with $\hat s=\pm0.2$ are shown in the $r$-$\phi$ plane for the choice of parameters $p=20$ and $e=2$ (so that $\chi_{\rm (max)}=2\pi/3$) and $\hat a=0.5$ in comparison with the geodesic of a spinless particle ($\hat s=0$). The values of both energy and angular momentum per unit mass are different depending on $\hat s$, whereas the minimum approach distance $r_{\rm(per)}\approx6.66667 M$ is the same for all cases. The deflection angle is $\delta\approx2.03215$ (i.e., $116.43382$ deg) for $\hat s=0$, $\delta\approx2.08603$ (i.e., $119.52073$ deg) for $\hat s=-0.2$, and $\delta\approx1.97828$ (i.e., $113.34690$ deg) for $\hat s=0.2$.
}
\label{fig:1}
\end{figure}

Figure \ref{fig:1} shows a typical hyperbolic-like orbit of a spinning particle with spin aligned along the positive $z$-axis and in the opposite direction compared with the corresponding geodesic orbit of a spinless particle.
The orbital parameters are chosen as $p=20$ and $e=2$, implying that the distance of minimum approach is $r_{\rm(per)}\approx6.66667 M$.
The trajectory of the spinning particle thus depends on the same parameters $p$ and $e$ as in the spinless case.
Once these parameters have been fixed, orbits with different values of $\hat s$ have the same closest approach distance, but different values of energy and angular momentum.  
On the contrary, setting a pair of values of ($\hat E$, $\hat J$) leads to a shift of the periastron due to spin, as shown below. 

                          
To first order in the rotation parameter $\hat a$ Eq. (\ref{dphidchi}) becomes
\begin{widetext}
\begin{eqnarray}
\frac{d\phi}{d\chi}
&=& \frac{1}{\sqrt{1-6u_p-2eu_p\cos \chi}}
-4\hat a \frac{\hat E_0u_p}{\hat J_0}\frac1{(1-6u_p-2eu_p\cos \chi)^{3/2}(1-2u_p-2eu_p\cos \chi)}\nonumber\\
&-&
\hat s \frac{\hat E_0u_p}{\hat J_0}\frac{e\cos\chi+3}{(1-6u_p-2eu_p\cos \chi)^{3/2}}\nonumber\\
&\equiv & \frac{d}{d\chi}\phi_0+\frac{d}{d\chi}\phi_{\hat a}+\frac{d}{d\chi}\phi_{\hat s}
\,,
\end{eqnarray}
\end{widetext}
where also terms of the order $\hat a \hat s$ have been neglected.
Integration then gives
\beq
\phi(\chi)=\phi_0(\chi)+\phi_{\hat a}(\chi)+\phi_{\hat s}(\chi)\,.
\eeq
The terms $\phi_0$ and $\phi_{\hat s}$ are given in Eq. (24) of Ref. \cite{Bini:2017ldh}, whereas the term linear in $\hat a$ is 
\begin{widetext}
\begin{eqnarray}
\phi_{\hat a}(\chi)&=&\hat a\frac{\hat E_0}{\hat J_0}\frac{\kappa}{\sqrt{eu_p}}\left[
\frac{E[(\pi-\chi)/2,\kappa]-E(\kappa)}{1-6u_p-2eu_p}
-\frac{\Pi[(\pi-\chi)/2,\nu,\kappa]-F[(\pi-\chi)/2,\kappa]}{1-2u_p+2eu_p}\right.\nonumber\\
&-&\left.
\frac{\kappa\sqrt{eu_p}}{1-6u_p-2eu_p}\frac{\sin\chi}{\sqrt{1-6u_p-2eu_p\cos\chi}}
\right]\,,\nonumber\\
\end{eqnarray}
\end{widetext}
with $\nu=4eu_p/(1-2u_p+2eu_p)$.

\subsection{Periastron shift}

A different (equivalent) parametrization of the orbit can be adopted in terms of energy $\hat E$ and angular momentum $\hat J$ instead of $p$ and $e$.
In this case the periastron distance depends on $\hat s$, and can be determined from the turning points for radial motion.

In the case of a spinless particle the radial and azimuthal equations can be conveniently written in the following factorized form in terms of the dimensionless inverse radial variable $u=M/r$ 
\begin{eqnarray}
\left(\frac{du}{d\tau}\right)^2&=&\frac{2\hat x^2}{M^2}u^4(u-u_{1\,(0)})(u-u_{2\,(0)})(u-u_{3\,(0)})\,,\nonumber\\
\frac{d\phi}{d\tau}&=&\frac{2\hat x}{M\hat a^2}u^2\frac{u_{4\,(0)}-u}{(u-u_+)(u-u_-)}\,,
\end{eqnarray}
which can be combined to yield
\begin{eqnarray} 
\label{dudphi_geo}
\left(\frac{du}{d\phi}\right)^2&=&\frac{\hat a^4}{2}\frac{(u-u_+)^2(u-u_-)^2}{(u_{4\,(0)}-u)^2}\times \nonumber\\
&& (u-u_{1\,(0)})(u-u_{2\,(0)})(u-u_{3\,(0)})\,.
\end{eqnarray}
Here $u_{1\,(0)}<u_{2\,(0)}<u_{3\,(0)}$ are the ordered roots of the equation
\beq
u^3-(\hat x^2+2\hat a\hat E\hat x+\hat a^2)\frac{u^2}{2\hat x^2}+\frac{u}{\hat x^2}+\frac{\hat E^2-1}{2\hat x^2}=0\,,
\eeq
with $\hat x=\hat J-\hat a\hat E$, whereas
\beq
u_\pm=\frac{M}{r_\pm}\,,\quad
u_{4\,(0)}=\frac{\hat J}{2\hat x}\,.
\eeq
For hyperbolic orbits we have $u_{1\,(0)}<0<u\leq u_{2\,(0)}<u_{3\,(0)}$, with $u_{2\,(0)}$ corresponding to the closest approach distance, i.e., $r_{\rm(per)}=M/u_{2\,(0)}$ \cite{Chandrasekhar:1985kt}.

The orbital equation for a spinning particle can be cast in a similar form as Eq. (\ref{dudphi_geo}), i.e.,
\begin{eqnarray}
\label{equdiphi}
\left(\frac{d u}{d \phi}\right)^2&=&
A\frac{\hat a^4}{2}(1-\hat sBu) \frac{(u-u_+)^2(u-u_-)^2}{(u-u_4)^2}\times
\nonumber\\
&& (u-u_1)(u-u_2)(u-u_3)
\,,
\end{eqnarray}
to first order in $\hat s$, where
\begin{eqnarray}
A&=&1+\frac{\hat s}{4\hat x^4}(\hat E\hat x+\hat a)[\hat a^3(3\hat E\hat x+\hat a)-2\hat x^2(2-\hat a^2)]\nonumber\\
&\equiv&1+\hat s A_{\hat s}
\,,\nonumber\\
B &=& -\frac{\hat a^2}{2\hat x^2}(\hat E\hat x+\hat a)\,,
\end{eqnarray}
with $u_i=u_{i\,(0)}+\hat s u_{i\,(1)}$.
The corrections to the geodesic values are
\begin{widetext}
\begin{eqnarray}
u_{1\,(1)}&=&-\frac{1-2u_{1\,(0)}+\hat a^2u_{1\,(0)}^2}{\hat x^2[{\hat x}(1-2u_{1\,(0)})+\hat a\hat E]}
\frac{\hat E({\hat E}^2-1+2u_{1\,(0)})-\hat x(\hat E\hat x+\hat a)u_{1\,(0)}^3}{(u_{1\,(0)}-u_{3\,(0)})(u_{1\,(0)}-u_{2\,(0)})}
\,,\nonumber\\
u_{2\,(1)}&=&-\frac{1-2u_{2\,(0)}+\hat a^2u_{2\,(0)}^2}{\hat x^2[{\hat x}(1-2u_{2\,(0)})+\hat a\hat E]}
\frac{\hat E({\hat E}^2-1+2u_{2\,(0)})-\hat x(\hat E\hat x+\hat a)u_{2\,(0)}^3}{(u_{2\,(0)}-u_{1\,(0)})(u_{2\,(0)}-u_{3\,(0)})}
\,,\nonumber\\
u_{3\,(1)}&=&-\frac{1-2u_{3\,(0)}+\hat a^2u_{3\,(0)}^2}{\hat x^2[{\hat x}(1-2u_{3\,(0)})+\hat a\hat E]}
\frac{\hat E({\hat E}^2-1+2u_{3\,(0)})-\hat x(\hat E\hat x+\hat a)u_{3\,(0)}^3}{(u_{3\,(0)}-u_{1\,(0)})(u_{3\,(0)}-u_{2\,(0)})}
\,,\nonumber\\
u_{4\,(1)}&=&-\frac{\hat a}{16\hat x^3}[(\hat E\hat x+\hat a)^3-8\hat E^2\hat x]
\,.
\end{eqnarray}
\end{widetext}
The periastron shift is then given by 
\beq
r_{\rm(per)}=\frac{M}{u_2}=\frac{M}{u_{2\,(0)}}\left(1-\hat s\frac{u_{2\,(1)}}{u_{2\,(0)}}\right)\,,
\eeq
to first order in spin.

Let us turn to the orbital equation (\ref{equdiphi}), which yields
\beq
\label{eqphidiu}
\frac{d \phi}{d u}=\pm\frac{\sqrt{2}}{\hat a^2\sqrt{A}}\frac{f(u)}{\sqrt{(u-u_1)(u_2-u)(u_3-u)}}\,,
\eeq
with 
\beq
f(u)=\left(1+\frac{B}{2}{\hat s}u\right)\frac{u_4-u}{(u-u_+)(u-u_-)}\,,
\eeq
where the $\pm$ sign should be chosen properly during the whole scattering process (with $u_1<0<u\leq u_2<u_3$), depending on the choice of initial conditions.
For instance, by choosing $\phi(u_2)=0$ at periastron, integration between $u_2$ and $u$ gives
\begin{widetext}
\begin{eqnarray}
\label{phidiusol2}
\phi(u)&=&\pm\frac{2\sqrt{2}}{\hat a^2(u_+-u_-)\sqrt{u_3-u_1}}\left\{
\left[1-\frac12\left(A_{\hat s}-Bu_+\right)\hat s\right]\frac{u_4-u_+}{u_1-u_+}\left(\Pi(\alpha,\beta_+,m)-\Pi(\beta_+,m)\right)\right.\nonumber\\
&&
-\left[1-\frac12\left(A_{\hat s}-Bu_-\right)\hat s\right]\frac{u_4-u_-}{u_1-u_-}\left(\Pi(\alpha,\beta_-,m)-\Pi(\beta_-,m)\right)\nonumber\\
&&
\left.
+\frac12B\hat s(u_+-u_-)\left[ K(m)-F(\alpha,m) \right]
\right\}
\,,
\end{eqnarray}
\end{widetext}
to first order in $\hat s$, where $\Pi(\alpha,0,m)=F(\alpha,m)$ and $\Pi(0,m)=K(m)$, and
\beq
m=\sqrt{\frac{u_2-u_1}{u_3-u_1}}\,,\quad
\alpha=\sqrt{\frac{u-u_1}{u_2-u_1}}\,,\quad
\beta_\pm=\frac{u_2-u_1}{u_\pm-u_1}\,.
\eeq
The total change in $\phi$ for the complete scattering process is then given by $2\phi(0)$ determined by Eq.~(\ref{phidiusol2}) with $\alpha=\alpha(0)=\sqrt{-u_1/(u_2-u_1)}$, so that the deflection angle is $\delta(\hat E,\hat J,\hat s)=2\phi(0)-\pi$.

\begin{figure}
\centering
\includegraphics[scale=0.4]{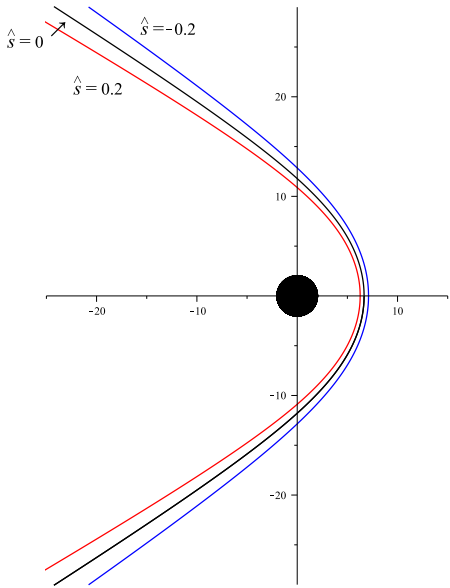}
\caption{The hyperbolic-like orbits of a spinning test particle with $\hat s=\pm 0.2$ are shown in the $r$-$\phi$ plane for the choice of parameters $\hat a=0.5$, $\hat E\approx1.08411$ and $\hat J\approx5.28307$ (so that $\hat x\approx4.74102$).
The reference geodesic is the same as in Fig. \ref{fig:1}.
The periastron is at  $u_2\approx0.14066$ (i.e., $r_{\rm(per)}\approx7.10950M$) for $\hat s=-0.2$, $u_2=0.15$ (i.e., $r_{\rm(per)}\approx6.66667 M$) for $\hat s=0$
 and $u_2\approx0.15934$ (i.e., $r_{\rm(per)}\approx6.27577M$) for $\hat s=0.2$.
The deflection angle is $\delta\approx2.03215$ (i.e., $116.43382$ deg) for $\hat s=0$, $\delta\approx1.95264$ (i.e., $111.87803$ deg) for $\hat s=-0.2$, and $\delta\approx2.10363$ (i.e., $120.52895$ deg) for $\hat s=0.2$.
}
\label{fig:2}
\end{figure}

A typical orbit using this parametrization is shown in Fig. \ref{fig:2}. 
The two branches (upper $+$, lower $-$) join at the periastron on the horizontal axis. 

                          
\subsection{Capture by the hole}

The condition for capture by the black hole is $u_2(\hat E, \hat J, \hat s) \ge u_3(\hat E, \hat J, \hat s)$, i.e., the cubic at the right hand side of Eq. (\ref{equdiphi}) has a double root at the critical point $u_2=u_3$ \cite{Misner:1974qy,Young:1976zz}.
Solving this condition for $\hat J$ then yields the critical value of the dimensionless total angular momentum ${\hat J}_{\rm crit}$ for capture, with associated dimensionless critical impact parameter $\hat b_{\rm crit}=b_{\rm crit}/M$ defined by\footnote{
The definition of the "impact parameter" should be viewed simply as a convenient rescaling of $J$. It can be formulated equivalently in terms of the particle's world line at infinity as $J=L+\hat E  s=m \sqrt{\hat E^2-1}  b +\hat E s$, (cf.\ Eqs.~(\ref{EandJ}) (as $r\to\infty$) and (\ref{ELs})), where L denotes the orbital angular momentum at infinity, see below.
}
\beq
\hat b_{\rm crit}=\frac{{\hat J}_{\rm crit}}{\sqrt{{\hat E}^2-1}}\,.
\eeq

The case of a spinless particle moving on a hyperbolic-like geodesic orbit in the equatorial plane is well known in both ultrarelativistic ($\hat E\gg1$) and non-relativistic ($\beta^2\equiv \hat E^2-1\ll1$) regimes (see, e.g., Ref. \cite{Frolov:1998wf}, Sec. 3.4.5, and references therein).
For example, the critical impact parameter for corotating orbits and $\hat a>0$ in the limit $\hat E\to\infty$ is given by (see Eq. (3.4.37) there)
\beq
\hat b_{(0)}=\hat a +8\cos^3\left(\frac13\arccos(-\hat a)\right)\,.
\eeq
For a spinning particle we find 
\begin{eqnarray}
{\hat b}_{\rm crit\,(0)}\big|_{\rm u.r.}&=&\hat b_{(0)}\left[1+\frac1{3\hat E^2}\left(1+\frac{\hat a}{\hat b_{(0)}}-{\hat s}k_{\hat s}\right)\right. \nonumber\\
&& \left.+O\left(\frac1{\hat E^4}\right)\right]\,,
\end{eqnarray}
where 
\begin{eqnarray}
k_{\hat s} &=&2\hat b_{(0)}\frac{(2w-2\hat a^2-\hat b_{(0)}^2)[6w-(\hat b_{(0)}+\hat a)^2]^3}{(\hat b_{(0)}+\hat a)^6(\hat b_{(0)}-\hat a)(\hat b_{(0)}-2\hat a)}\times \nonumber\\
&&
\frac{w-\hat a(\hat b_{(0)}+\hat a)}{2w-\hat b_{(0)}(\hat b_{(0)}+\hat a)}
\,,
\end{eqnarray}
with $w=\sqrt{3(\hat b_{(0)}^2-\hat a^2)}$.
To linear order in the rotational parameter $\hat a$ (also neglecting terms $\hat a\hat s$) the previous expression becomes
\begin{eqnarray}
{\hat b}_{\rm crit\,(0)}\big|_{\rm u.r.}&\simeq & 3\sqrt{3}\left(1-\frac{2}{3\sqrt{3}}\hat a\right)\times \nonumber\\
&& \left[1+\frac{1}{3\hat E^2}\left(1+\frac{1}{3\sqrt{3}}(\hat a+\hat s)\right)\right. \nonumber\\
&& \left. +O\left(\frac1{\hat E^4}\right)\right]\,.
\end{eqnarray}

In the non-relativistic limit the critical impact parameter for corotating geodesic orbits and $\hat a>0$ is given by (see Eq. (3.4.33) of Ref. \cite{Frolov:1998wf})
\beq
\hat b_{(0)}=\frac{2}{\beta}\left(1+\sqrt{1-\hat a}\right)\,.
\eeq
For a spinning particle we find 
\beq
{\hat b}_{\rm crit\,(0)}\big|_{\rm n.r.}=\hat b_{(0)}\left[1+\hat s\frac{\sqrt{1-\hat a}}{2(1+\sqrt{1-a})^2}\right]+O(\beta^0)\,,
\eeq
which in the limit of small rotation parameter reduces to  
\beq
{\hat b}_{\rm crit\,(0)}\big|_{\rm n.r.}\simeq\frac4{\beta}\left(1-\frac14\hat a\right)\left(1+\frac18\hat s\right)+O(\beta^0)\,.
\eeq

\section{Including quadratic-in-spin corrections}

In the previous section we have seen how (linear) corrections due to spin affect the geodesic scattering angle.
According to the positive/negative sign of the spin the scattering angle is larger/smaller than the corresponding geodesic value. It is interesting to consider  quadratic-in-spin corrections too, not sensitive to the sign of the spin. This requires generalizing the above discussion to include quadrupolar terms in the MPD equations.
The presence of a completely general quadrupolar modification to the MPD equations and subsequent derivation of the scattering angle is beyond the scope of this work. We will limit here our analysis to the simpler case of a \lq\lq black-hole-like" extended body only, i.e., to the case of a body with special spin-squared quadrupolar structure, as specified below.

Let us recall that the MPD equations for bodies with general quadrupolar structure read \cite{Steinhoff:2012rw,Bini:2013nw,Hinderer:2013uwa,Bini:2013uwa}
\begin{eqnarray}
\label{MPDp}
\frac{DP_\mu}{d\tau}&=&-\frac{1}{2}R_{\mu\nu\alpha\beta}U^\nu S^{\alpha\beta}-\frac{1}{6}\nabla_\mu R_{\nu\rho\alpha\beta}J^{\nu\rho\alpha\beta}\phantom{\bigg|}
\,,\nonumber\\
\label{MPDS}
\frac{DS^{\mu\nu}}{d\tau}&=&2P^{[\mu}U^{\nu]}+\frac{4}{3}R^{[\mu}{}_{\rho\alpha\beta}J^{\nu]\rho\alpha\beta}\,,
\end{eqnarray}
where $J^{\mu\nu\alpha\beta}$ denotes the relativistic quadrupole tensor, which has all the algebraic symmetries of a Riemann tensor.
When specialized to the ``spin-squared'' case it reads
\beq\label{BHJ}
J^{\mu\nu\alpha\beta}=\frac{3P\cdot U}{P^2}u^{[\mu}S^{\nu]\rho} u^{[\alpha}S^{\beta]}{}_\rho\,,
\eeq
corresponding to an extended body with a \lq\lq black-hole-like quadrupole" tensor\footnote{
A more general expression uses a polarizability parameter $C_Q$ as a proportionality constant in the rhs of Eq. (\ref{BHJ}), related to the shape of the test body, and can be found for example in \cite{Hinderer:2013uwa}; see also Ref. \cite{Bini:2015zya}, Eqs. (2.6) and (2.7) where $C_Q=1$ to compare with the present case.
}.  
As before, it is useful to trade $S^{\mu\nu}$ for the spin vector $S(u)^\mu$ obtained by the spatial duality operation of Eq. (\ref{svec}).
The  relationship between $U$ and $P=mu$ then follows from differentiating along $U$ the spin supplementary conditions and using Eqs. (\ref{MPDp}). 
For the \lq\lq black-hole-like" extended body a direct calculation shows  that $P$ and $U$ remain aligned, i.e., $u^\mu=U^\mu+O(S^3)$ (see, e.g., Eq. 4.21 of Ref. \cite{Bini:2015zya} with $C_Q=1$).
However, the mass $m=||P||$ is no longer a conserved quantity, and is given by
\begin{eqnarray}
m&=&m_0+\frac{1}{6}R_{\mu\nu\alpha\beta}J^{\mu\nu\alpha\beta}\nonumber\\
&=& m_0-\frac{1}{2m_0}E(U)_{\nu \beta} S(u)^\nu   S(u)^\beta+O(S^3)\,,
\end{eqnarray}
where $E(U)_{\nu\beta}=R_{\mu\nu\alpha\beta}U^\mu U^\alpha$ is the electric part of the Riemann tensor with respect to $U$ and $m_0$ denotes the (constant) \lq\lq bare'' mass of the body.  

\begin{figure}
\centering
\includegraphics[scale=0.35]{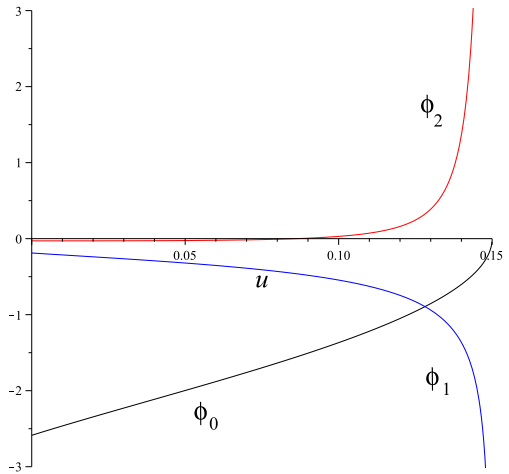}
\caption{
The behavior of each individual contribution $\phi_0$, $\phi_1$ and $\phi_2$ to the azimuthal angle $\phi$ is shown as a function of $u$ for the same choice of parameters as in Fig. \ref{fig:2} (corresponding to the lower branch there), namely $\hat a=0.5$, $\hat E\approx1.08411$ and $\hat J\approx5.28307$. The spin dependent term are both indefinitely growing as the minimum approach distance is reached, but with opposite behavior. With these values of parameters one can compute the minimum approach distance and the result is
$u_2\approx0.15 +0.0467\hat s  +  0.0124\hat s^2$.
}
\label{fig:3}
\end{figure}

For equatorial motion with aligned spin (see Eq. (\ref{SUdef})) the spin evolution equations are satisfied with constant signed spin magnitude $s$, and the orbital motion is fully determined by the conservation of energy and total angular momentum still written in the form (\ref{totalenergy}).
The orbital equation then reads
\begin{eqnarray}
\label{du_dphi}
\left(\frac{du}{d\phi}\right)^2 &=& {\mathcal V}(u; \hat E,\hat L, \hat s)+O(\hat s^3)\nonumber\\
&=& {\mathcal V}_0(u; \hat E,\hat L)+ \hat s {\mathcal V}_1(u; \hat E,\hat L)\nonumber\\
&& +\hat s^2 {\mathcal V}_2(u; \hat E,\hat L) +O(\hat s^3)\,,
\end{eqnarray}
with
\begin{widetext}
\begin{eqnarray} 
{\mathcal V}(u; \hat E,\hat L, \hat s)&= &\frac{\hat\Delta^2}{\hat w^2}\Bigg\{(\hat E^2-1)(1+\hat a^2u^2)
-\hat L^2u^2+(u+\hat y^2u^3)\Big(2+3\hat\Delta u^2\hat s^2\Big)
\nonumber\\&+&
\frac{\hat\Delta u^3}{\hat w}\left[-(\hat E\hat y+\hat a)\left[2\hat y\hat s+\left(2\hat E+\frac{3\hat a\hat y^2u^3}{\hat w}\right)\hat s^2\right]  +\hat y^3u^3\hat s^2\right]\Bigg\}\,,
\end{eqnarray}
\end{widetext}
where
\beq
\hat\Delta=1-2u+\hat a^2u^2\,,\quad 
\hat y=\hat L-\hat a\hat E\,,\quad  
\hat w=\hat L-2\hat yu\,.
\eeq
The dimensionless energy and angular momentum $\hat E$ and $\hat J$ as well as the test-body's orbital angular momentum $\hat L$ (at infinity) are now defined by using the conserved bare mass $m_0$, namely
\begin{eqnarray}
\label{ELs}
\hat E&=&\frac{E}{m_0},\qquad \hat L=\hat J-\hat E \hat s,\nonumber\\
\hat L&=&\frac{L}{m_0M},\qquad\hat s=\frac{s}{m_0M}\,.
\end{eqnarray}
Equivalently, one may take as parameters $\hat E$, $\hat s$ and $\hat J$ (in place of $\hat L$) so that the above equation (formally) becomes
\begin{eqnarray}
\label{du_dphi2}
\left(\frac{du}{d\phi}\right)^2 &=& {\mathcal W}(u; \hat E,\hat J, \hat s)+O(\hat s^3)\nonumber\\
&=&  {\mathcal W}_0(u; \hat E,\hat J)+ \hat s {\mathcal W}_1(u; \hat E,\hat J)\nonumber\\
&& +\hat s^2 {\mathcal W}_2(u; \hat E,\hat J) +O(\hat s^3)\,.
\end{eqnarray}
The solution of Eq. (\ref{du_dphi2}) can then be formally written as
\beq
\phi(u)=\phi_0(u; \hat E,\hat J)+\hat s \phi_1(u; \hat E,\hat J)+\hat s^2 \phi_2(u; \hat E,\hat J)+O(\hat s^3)\,.
\eeq
It turns out that the  $\hat s^2$ corrections are  more involved with respect to the linear ones and the discussion best proceeds by working numerically, or by using approximation schemes such as the post-Newtonian and post-Minkowskian ones. Fig. \ref{fig:2} can be redone then to include the $\hat s^2$ modifications to the orbit; however, with the same choice of parameters as in Fig. \ref{fig:2} the differences cannot be visually appreciated. It is interesting to study the behavior of the various separated contributions $\phi_n(u; \hat E,\hat J)$ ($n=0,1,2$) as shown in Fig. \ref{fig:3}. At the minimum approach distance the geodesic term $\phi_0$ vanishes (as the full quantity $\phi(u)$), while both the linear and quadratic corrections diverge, one positively ($\phi_2$) and the other negatively ($\phi_1$), leading to compensating effects.

                          
\section{Concluding remarks}

We have studied here the hyperbolic scattering of a two body system consisting of a rotating (Kerr) black hole and a spinning test body described according to the Mathisson-Papapetrou-Dixon model. The motion has been assumed to be equatorial, i.e., confined on the reflection-symmetry plane of the Kerr spacetime, with the spin of the body aligned/anti-aligned with that of the hole, implying meaningful simplifications and allowing for a handy treatment.
Backreaction effects on the background metric are neglected because of the smallness condition of the characteristic spin length scale with respect to the background curvature length scale implicit in the MPD model. 

Taking advantage of the existence of conserved quantities related to Killing symmetries, we have been able to analytically solve the \lq\lq linearized in spin'' equations of motion in terms of elliptic integrals.
We have computed the scattering angle to linear order in spin by using two equivalent parametrizations of the orbit: in terms of either eccentricity and (inverse) semi-latus rectum, or total energy and angular momentum (which are gauge-invariant variables). 
In the former case orbits with different values of the spin magnitude have the same closest approach distance, but different values of energy and angular momentum.  
On the contrary, a given pair of values of energy and angular momentum (the same for both geodesic and accelerated orbits) leads to a shift of the periastron due to spin. In this case the geodesic value of the scattering angle is modified in such a way that it increases/decreases according to the spin orientation being aligned/anti-aligned with the black hole rotation axis, respectively.

Finally, we have shown how to generalize the present analysis to the case of extended bodies with quadrupolar structure, even if in the simplified situation of spin-squared quadrupole interactions for a \lq\lq black-hole-like'' body.
The presence of a quadrupolar force term in the equations of motion complicates matters, so that obtaining a closed form expression for the orbit is not as simple as in the linear-in-spin case. We leave to a forthcoming paper the discussion of more general quadrupolar situations, which plays a role in the current analysis of binary dynamics.

\section*{Ackwnoledgments}
J.V. thanks J. Steinhhoff  for useful discussions. D.B. thanks T. Damour for many useful discussions and the Albert Einstein Institute for the warm hospitality at the beginning of the present project.

\end{document}